\documentclass[9pt,twocolumn,twoside]{optica}
\setboolean{shortarticle}{true}
\setboolean{minireview}{false}
\usepackage{graphicx,subfigure,amssymb,mathtools,bm,nicefrac,blindtext} 

\newcommand{\openone}{\leavevmode\hbox{\small1\normalsize\kern-.33em1}} 

\newcommand{\op}[1]{\hat{#1}}

\title{Quantum metrology at the limit with extremal Majorana constellations}

\author[1]{F.~Bouchard}
\author[2]{P.~de~la~Hoz}
\author[3]{G.~Bj\"{o}rk} 
\author[1,4]{R.~W.~Boyd}
\author[5]{M.~Grassl} 
\author[6]{Z.~Hradil} 
\author[1,7,*]{E.~Karimi}
\author[8]{A.~B.~Klimov} 
\author[5,1]{G.~Leuchs}
\author[6]{J.~\v{R}eh\'{a}\v{c}ek} 
\author[2,5]{L.~L.~S\'anchez-Soto} 

\affil[1]{Department of Physics, University of Ottawa, 
150 Louis Pasteur, Ottawa, Ontario, K1N 6N5 Canada}
\affil[2]{Department  of Optics, Faculty of Physics, 
Universidad Complutense, 28040~Madrid, Spain}
\affil[3]{Department of Applied Physics, 
Royal Institute of Technology (KTH), AlbaNova, 
SE-106 91 Stockholm, Sweden}
\affil[4]{Institute of Optics, University of Rochester, 
Rochester,  New York, 14627, USA}
\affil[5]{Max Planck Institute for the Science of Light,
  Staudtstra\ss e 2, 91058 Erlangen, Germany} 
\affil[6]{Department of Optics,  Palack\'y  University, 
17.~listopadu 12, 771 46 Olomouc,  Czech Republic}
\affil[7]{Department of Physics, 
Institute for Advanced Studies in Basic Sciences, 
45137-66731 Zanjan, Iran}
\affil[8]{Department of Physics, 
Universidad de Guadalajara, 44420~Guadalajara, 
Jalisco, Mexico}
\affil[*]{Corresponding author: ekarimi@uottawa.ca}

\dates{Compiled \today}

\ociscodes{(270.0270) Quantum optics; 
 (270.5585)   Quantum information and processing; 
(120.3940)   Metrology.}

\doi{\url{http://dx.doi.org/10.1364/aop.XX.XXXXXX}}

\begin{abstract}
  Quantum metrology allows for a tremendous boost in the accuracy of
  measurement of diverse physical parameters. The estimation of a
  rotation constitutes a remarkable example of this quantum-enhanced
  precision. {The recently introduced Kings of Quantumness are
    especially germane for this task when the rotation axis is
    unknown, as they have a sensitivity independent of that axis and
    they achieve a Heisenberg-limit scaling}.  Here, we report the
  experimental realization of these states by generating up to
  21-dimensional orbital angular momentum states of single photons,
  and confirm {their high metrological abilities.}
\end{abstract}

\setboolean{displaycopyright}{true}

\begin{document}

\maketitle
\thispagestyle{fancy}
\ifthenelse{\boolean{shortarticle}}{\abscontent}{}

The conventional description of the quantum world involves a key
mathematical object---the quantum state---that conveys complete
information about the system under study: once it is known, the
probabilities of the outcomes of any measurement can be
predicted. This statistical description entails counterintuitive
effects that have prompted several notions of quantumness, yet no
single one captures the whole breadth of the physics.  

There are, however, instances of quantum states that behave in an
almost classical way. The paradigm of such a behavior is that of
coherent states of light~\cite{Glauber:1963aa}: they are as much
 localized as possible in phase space, a property that is preserved
under free evolution.

The concept of coherent states has been extended to other physical
systems~\cite{Perelomov:1986ly}. The case of a spin is of paramount
importance. The corresponding spin coherent states have minimal
uncertainty and they are conserved under rotations. So, in the usual
way of speaking, they mimic a classical angular momentum as much as
possible. One could rightly wonder what kind of state might serve as
the opposite of a coherent state. The answer will depend on the ways
to formalize the idea of being ``the
opposite''~\cite{Zimba:2006fk}. Here, we take advantage of the
Majorana representation, which maps a pure spin $S$ into $2S$ points
on the Bloch sphere~\cite{Majorana:1932ul}.  

It turns out that the Majorana representation of a coherent state
consists of a single point (with multiplicity $2S$). At the opposite
extreme, we can imagine states whose Majorana representations are
spread uniformly over the sphere. The resulting states are
precisely the Kings of Quantumness~\cite{Bjork:2015aa,Bjork:2015ab}.
With such symmetric spreadings, the constellations essentialy map onto
themselves for relatively small rotations around arbitrary axes. This
means that they  resolve rotations around any axis approximately
equally well.  We emphasize that the problem of estimating a rotation is of utmost
interest in magnetometry~\cite{Wasilewski:2010aa,Sewell:2012aa,Muessel:2014aa}, 
polarimetry~\cite{Meyer:2001aa,DAmbrosio:2013aa}, and metrology in
general~\cite{Rozema:2014aa}.  In this work, we experimentally
demonstrate the generation of these states and certify their potential
for quantum metrology~\cite{Giovannetti:2011aa}.

Let us first set the stage for our experiment. We consider a system
that can be described in terms of two independent bosonic modes, with
creation operators $\op{a}_{\alpha}^\dag$, with
$\alpha \in \{ +, -\}$.  This encompasses many different instances,
such as strongly correlated systems, light polarization, Bose-Einstein
condensates, and Gaussian-Schell beams, to mention only but a
few~\cite{Chaturvedi:2006vn}.  The Stokes operators for these two-mode
systems can be compactly expressed as~\cite{Luis:2000ys}
$\op{\mathbf{S}} = \tfrac{1}{2}  \op{a}_{\alpha}^{\dagger} \bm{\sigma}_{\!\alpha \beta} \,
\op{a}_{\beta}$,  where $\bm{\sigma}$ denote the Pauli matrices and summation over
 repeated indices is assumed.  One can verify that
 $\op{\mathbf{S}}^{2} = \op{S}_{0} ( \op{S}_{0} + \openone)$, with 
$\op{S}_{0} = \op{N}/2$ and $\op{N} = \op{a}_{\alpha}^{\dagger}
\delta_{\alpha \beta} \, \op{a}_{\beta} = \op{N}_{+} + \op{N}_{-}$
being the total number of  excitations. 

From now on, we restrict our attention to the case where $N$ is fixed.
{This corresponds to working in a $(2S+1)$-dimensional Hilbert space
$\mathcal{H}_{S}$ of spin $S$ (with $N = 2S$).} This space
$\mathcal{H}_{S}$ is spanned by the Dicke basis $|S, m \rangle$,
wherein the action of $\op{\mathbf{S}}$ operators is the standard for
an angular momentum. Sometimes, it is preferable to use the two-mode
Fock basis $|N_{+}, N_{-} \rangle$;  related to the Dicke basis
by  $N_{+} = S + m$ and $N_{-} = S - m$.

\begin{figure}
  \centerline{\includegraphics[width=.90\columnwidth]{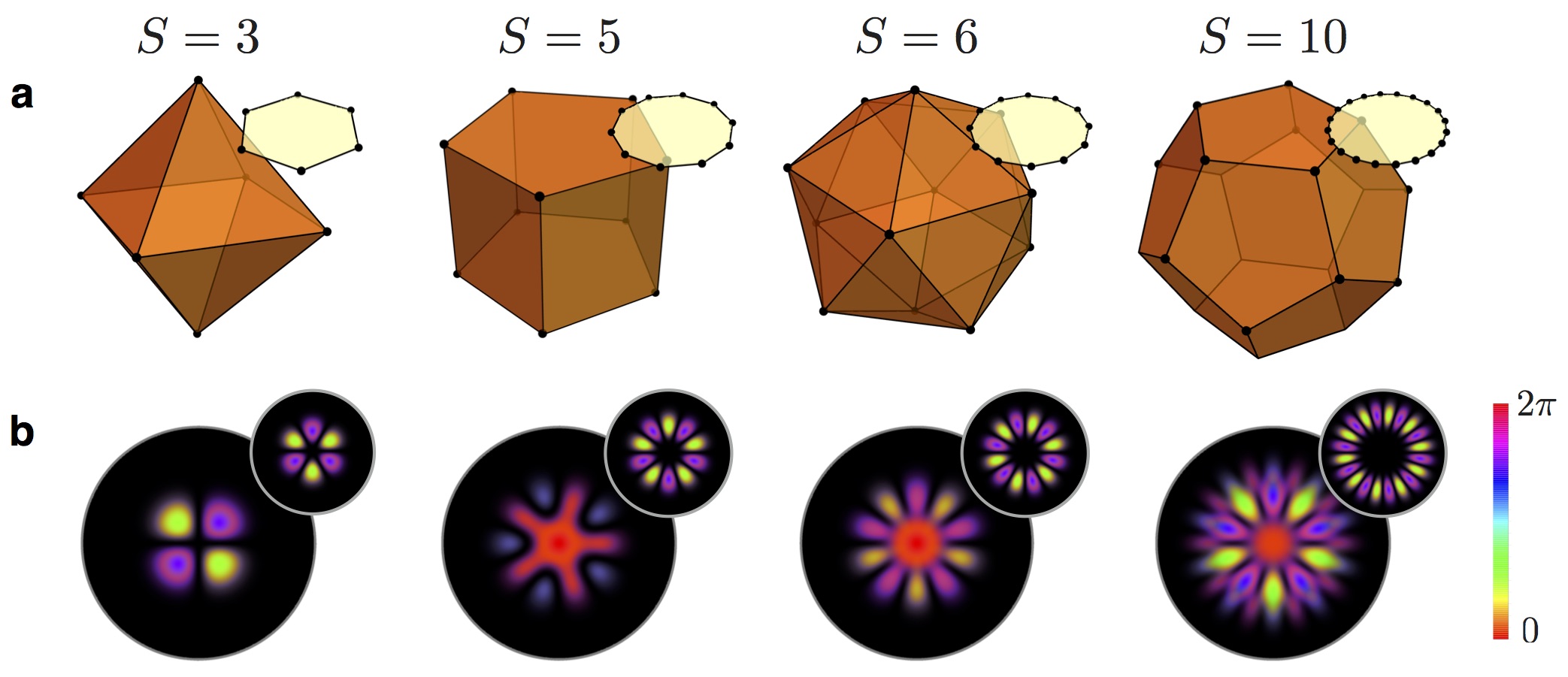}}
  \caption{(a) The Majorana constellations in the Bloch sphere for the Kings
    (orange) and the NOON states (yellow) corresponding to spin
    $S = 3, 5, 6$, and $10$. (b) The Laguerre-Gauss
    representation of the same Kings and NOON states, shown in
    (a), where the azimuthal index $\ell$ corresponds to $m$ in
    the Dicke basis. We consider the fundamental radial mode; i.e.,
    $p=0$, where $p$ is the radial index of the Laguerre-Gauss modes.}
  \label{fig:const}
\end{figure}

Spin coherent states are constructed much in the same way as in the
canonical case~\cite{Perelomov:1986ly}: they are displaced versions of
the north pole of the Bloch unit sphere $\mathcal{S}_{2}$. If
$\mathbf{n}$ is a unit vector in the direction of the spherical angles
($\theta , \phi$), they can be defined as
$| \mathbf{n} \rangle = e^{i \phi \op{S}_{z}} e^{i \theta \op{S}_{y}}
|S, S \rangle$.  They are not orthogonal, but one can still decompose
an arbitrary state $|\Psi \rangle$ using this overcomplete set.  The
associated coherent-state wave function is
$\Psi (\mathbf{n}) = \langle \mathbf{n} | \Psi \rangle$, and the
corresponding probability distribution,
$Q (\mathbf{n}) = |\Psi (\mathbf{n})|^{2}$, is nothing but the Husimi
function.

The wave function $\Psi (\mathbf{n}) $ can be expanded in
terms of the Dicke basis $| {S, m} \rangle$. If the
corresponding coefficients are 
$\Psi_{m} = \langle {S, m} | \Psi \rangle$, we obtain
$ \Psi ( \mathbf{n} ) = (1 + |z|^{2})^{- S} \sum_{m=-S}^{S} c_{m}
\Psi_{m} \, z^{S+m}$, where $c_{m} = \sqrt{(2S)!/[(S-m)! (S+m)!]}$ and
$z = \tan (\theta/2) e^{-i \phi} $ is the complex number derived by
the stereographic projection of $(\theta , \phi)$. Apart from the
unessential positive prefactor, this is a polynomial of order $2S$;
thus, $| \Psi \rangle$ is determined by the set $\{ z_{i} \}$ of the
$2S$ complex zeros of $\Psi ( \mathbf{n} )$. These zeros, which are
also the zeros of $Q (\mathbf{n})$, specify the so-called
constellation by an inverse stereographic map of
$\{ z_{i}\} \mapsto ( \theta_{i}, \phi_{i} )$.

Since the spherical harmonics $Y_{Kq} (\mathbf{n}$ are a complete set
of orthonormal functions on $\mathcal{S}_{2}$, they may be used to
expand the Husimi function $Q(\mathbf{n})$. {The resulting
  coefficients, $\varrho_{Kq}$, are nothing but the standard state
  multipoles~\cite{Blum:1981rb} and there are $2S+1$ of them (see
  Supplemental material)}. The monopole is trivial, as it is just a
constant term. The dipole indicates the position of the state in the
Bloch sphere. When it vanishes, the state has vanishing (first-order)
polarization and points nowhere in the mean. If the quadrupole also
vanishes, the variance of the state is uniform; i.e., no directional
signature can be observed in its second-order fluctuations and we say
that it is second-order unpolarized.  Similar interpretation holds for
higher-order multipoles. One can also look at these multipoles as the
$K$th directional moments of the state constellation and, therefore,
these terms resolve progressively finer angular features.

\begin{figure}
  \centerline{\includegraphics[width=0.90\columnwidth]{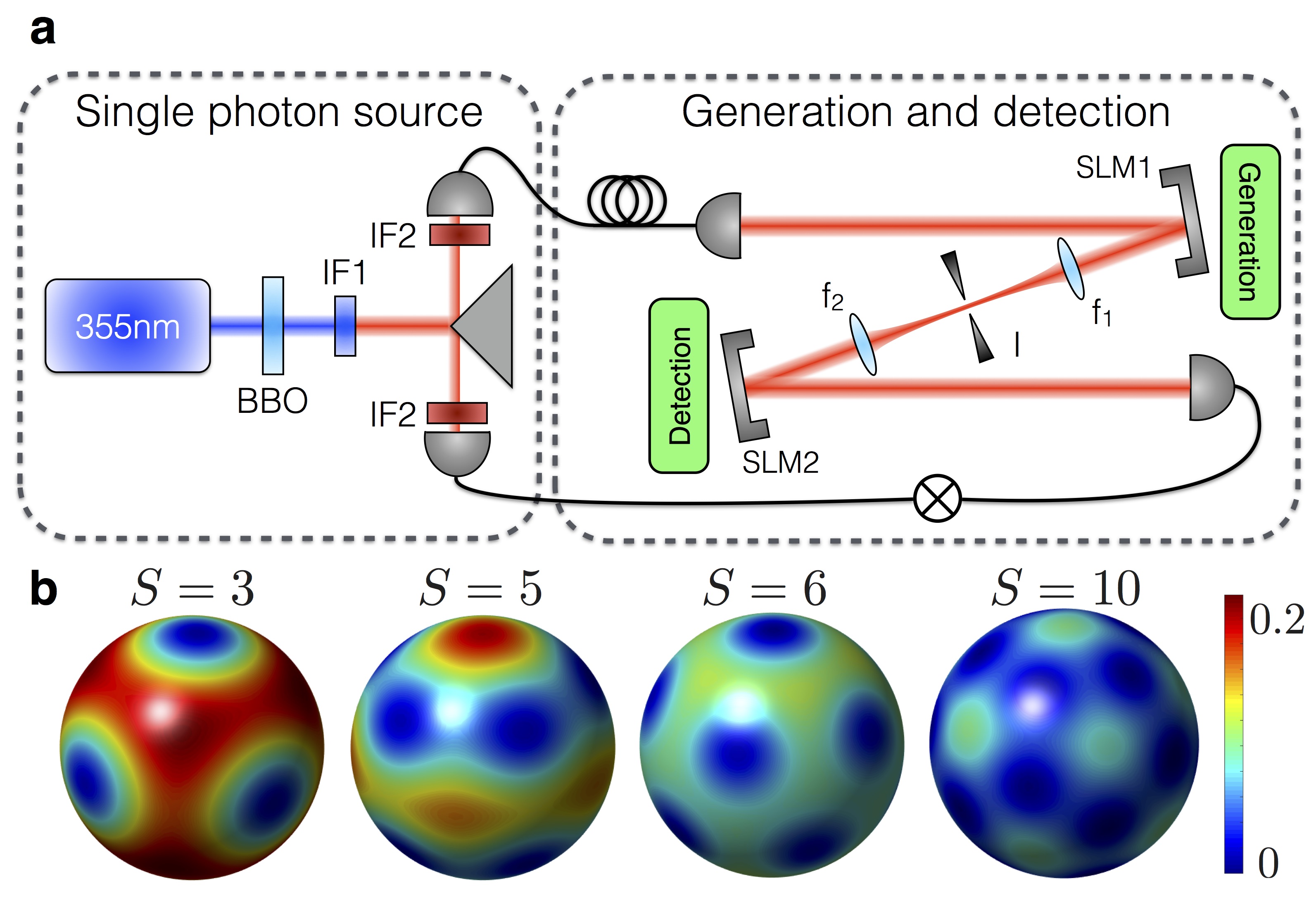}}
  \caption{(a) Sketch of the experimental setup and (b) density plots
    of the experimentally reconstructed Husimi $Q$ functions for the
    same King states as in Fig.~\ref{fig:const}. The fidelities of these
    reconstructed states are (from left to right) 0.94, 0.87, 0.91,
    and 0.93. The differences with the theoretical $Q$ functions
    cannot be visually noticed.}
  \label{fig:setup}
\end{figure}

The quantity $\sum_{q} | \varrho_{Kq} |^{2} $ gauges the
overlap of the state with the $K$th multipole pattern.  It seems thus
suitable to look at the cumulative distribution~\cite{Hoz:2013om}
$\mathcal{A}_{M} = \sum_{K= 1}^{M} \sum_{q=- K}^{K} | \varrho_{Kq}
|^{2}$, which concisely condenses the state angular capacity up to
order $M$ ($1 \le M \le 2S$). Observe that the monopole is omitted, as
it is just a constant term.

\begin{figure*}[t]
  \centerline{\includegraphics[width=1.50\columnwidth]{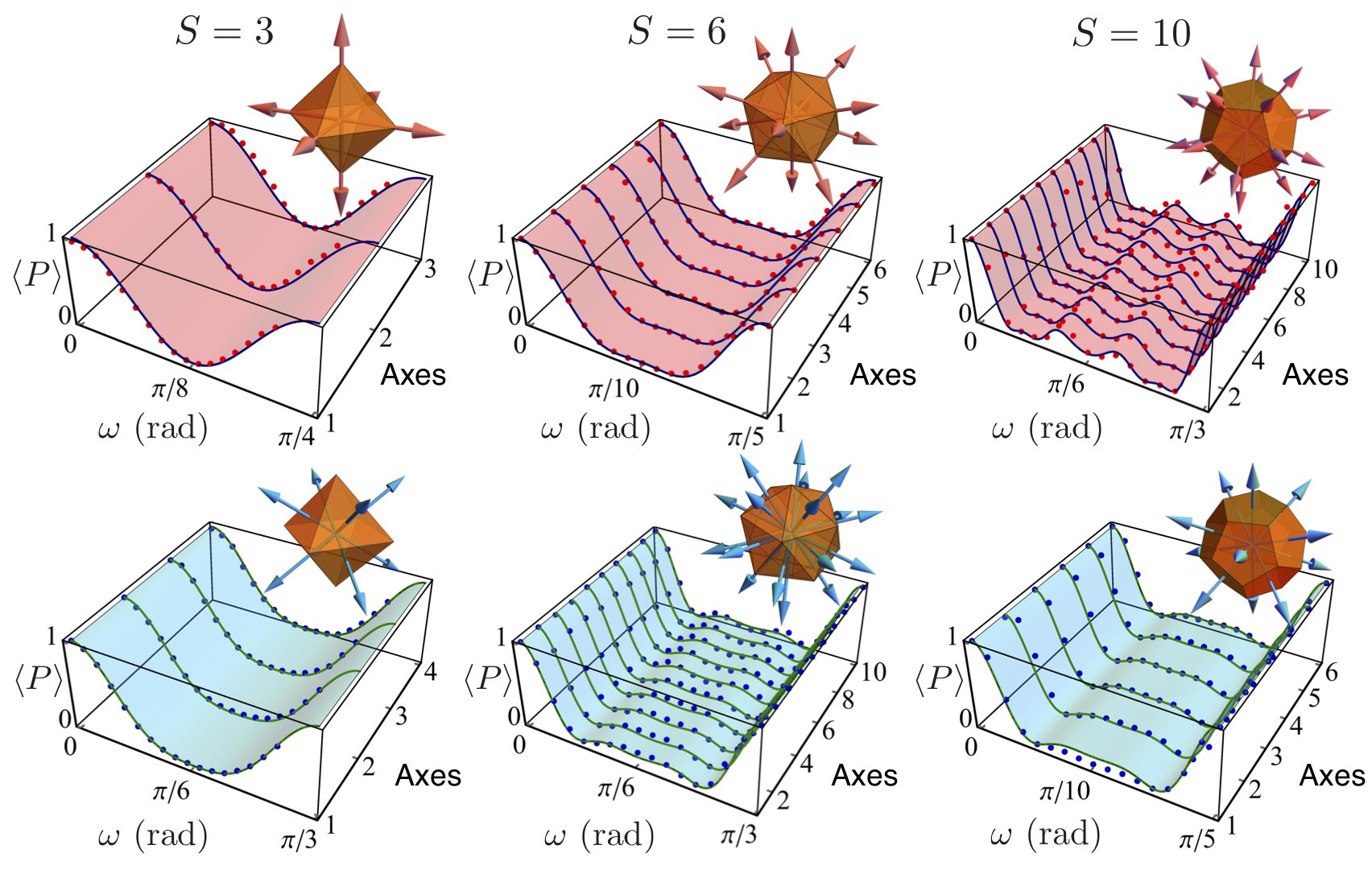}}
  \caption{ Experimental results of the projection of the $S=3,6$ and
    $10$ (first, second, and third column, respectively) Kings states,
    $|\Psi^{(S)} \rangle$, onto themselves after a rotation of
    $\omega$ around the axis $\bm{\mathfrak{u}}$,
    $ \hat{R} (\omega, \bm{\mathfrak{u}} )$; i.e.,
    $\langle \hat{P} \rangle= | \langle \Psi ^{(S)}| \hat{R} (\omega,
    \bm{\mathfrak{u}}) | \Psi^{(S)} \rangle |^2$. The axes are
    presented graphically along with the associated
    constellations. The first row corresponds to rotations along the
    axes passing through the Majorana points (pink arrows) and the
    second row corresponds to rotations along the axes normal to the
    facets of the constellations (blue arrows). The experimental
    results (red and blue dots) are shown along with the theoretical
    results (blue and green curves) for all rotation axes.}
  \label{fig:sensitivity}
\end{figure*}

The spin coherent states $| \mathbf{n} \rangle$ have remarkably simple
constellations, just the point $-\mathbf{n}$, and maximize
$\mathcal{A}_{M}$ for all orders $M$, confirming yet from another
perspective the outstanding properties of these
states~\cite{Bjork:2015aa}.  

In contradistinction, the Kings are those pure states that make
$\mathcal{A}_{M} \equiv 0$ for the highest possible value of $M$.
This means that they convey the relevant information in higher-order
fluctuations. The search for these states has been systematically
undertaken recently in Ref.~\cite{Bjork:2015aa}, where the interested
reader can check the details (see also Supplemental Material,
{where one can find the nonzero components $ \Psi_{m}$ of the
  Kings}).  The resulting Majorana constellations for some values of
$S$ are depicted in Fig.~\ref{fig:const}. For $S=3$, the constellation
is a regular octahedron and the state is third-order unpolarized
($M=3$). For $S=5$, it consists of two pentagons. For $S=6$ we have
the icosahedron, and the corresponding state is fifth-order
unpolarized.  For $S=10$ we have a slightly stretched dodecahedron
(i.e., the four pentagonal rings that define its vertices are
displaced against the pole) and it is fifth-order
unpolarized. {As we can appreciate, the Kings have the points
  very symmetrically placed on the unit sphere, so their
  constellations possess many axes along which they return to
  themselves after a rotation.} Consequently, they can resolve
relatively small angles around a large number of axes.

Other states with a high degree of angular resolution are the NOON
states, given by
$| \mathrm{NOON} \rangle = \tfrac{1}{\sqrt{2}} ( |N, 0 \rangle - |0, N
\rangle)$ in the two-mode Fock basis and
$ \tfrac{1}{\sqrt{2}} (|S, S \rangle - |S, -S \rangle)$ in the Dicke
basis. As shown in Fig.~\ref{fig:const} their Majorana constellation
consists of $2S$ equidistantly placed points around the equator of
$\mathcal{S}_2$.  A rotation around the $z$ axis of angle $\pi/(2 S)$
makes $| \mathrm{NOON} \rangle$ orthogonal to itself, whereas for
$\pi/S$ it returns to itself. This nicely supports the ability of NOON
states to detect small rotations.

To compare the performance of these two classes of states, let us
assume we have to estimate a rotation
$R ( \omega, \bm{\mathfrak{u}} )$ of angle $\omega$ around an axis
$\bm{\mathfrak{u}}$ of spherical angles $(\Theta, \Phi )$.  We
consider only small rotations and take
the measurement to be a projection of the rotated state onto the
original one; i.e., it can be represented by
$\op{P} = | \Psi \rangle \langle \Psi |$. 
As discussed in the
Supplemental material, the respective sensitivities ({defined as
  the ratio $ \Delta \omega = \Delta \op{P}/ | \partial \langle \op{P}
  \rangle/\partial \omega |$,  the variance being $\Delta \op{P} = 
[ \langle \op{P}^{2}\rangle - \langle \op{P} \rangle^2]^{1/2}$})  are
\begin{eqnarray}
  \label{eq:senKN}
  \Delta \omega_{\mathrm{Kings}} &  =  &
\frac{\sqrt{3}}{2}
\frac{1}{\sqrt{S(S+1)}} \, ,
\nonumber \\
  \Delta \omega_{\mathrm{NOON}}  & =  & 
 \frac{1}{\sqrt{2}} 
 \frac{1}{\sqrt{2 S^{2} \cos^{2} \Theta + S \sin^{2} \Theta}} \,.
\end{eqnarray}
 
The sensitivity of the Kings is completely independent of the rotation
axis and with a Heisenberg-limit scaling $1/S$ for large $S$. For the
NOON states, the sensitivity scales as $1/S$  when $\Theta = 0$, but
can be as bad as $1/\sqrt{S}$ when $\Theta = \pi/2$. In short, it is
essential to know the rotation axis to ensure that the NOON state is
aligned to achieve its best sensitivity. 

We stress that the measurement scheme for $\Delta \omega$
involves only second-order moments of $\op{\mathbf{S}}$. Given their
properties, one could expect that detecting higher-order moments will
bring out even more advantages of the Kings. 
 
To check these issues we have generated these extremal states for the
cases of $S=3, 5, 6$ and 10, {using orbital angular momentum
  (OAM) states of single photons~\cite{Allen:2003aa}, which has
  already proven fruitful in quantum metrology~\cite{Simon:2017aa}.
  Working at the single-photon regime is not essential, but it
  highlights the potential implications for quantum information
  processing~\cite{Leach:2002aa}. Therefore, the index $m$ in the
  Dicke basis is identified with the OAM eigenvalue $\ell$ of a single
  photon along its propagation direction. In general, there exist many
  families of optical modes carrying OAM, but we choose the
  Laguerre-Gauss basis LG$_{\ell, p}$, where $p$ is the radial index.
  Since the radial profile is irrelevant to the experimental
  realization of the Kings states, for the sake of simplicity, we
  always set the radial index to its fundamental value, i. e. $p=0$.
  The resulting transverse profiles of both the Kings and the NOON
  states are as in Fig.~\ref{fig:const}(b).}

We experimentally create {the Kings by means of spontaneous
  parametric downconversion.} A sketch of the experimental setup is
shown in Fig.~\ref{fig:setup}(a). A quasi-continuous wave UV laser
operating with a repetition rate of 100~MHz and an average power of
150~mW at a wavelength of 355~nm is used to pump a type-I
$\beta$-barium borate crystal.  The single photons, signal and idler,
are subsequently coupled to single mode fibres to filter their spatial
mode. One of the photons, the idler, is used as a trigger. The other
photon, the signal, is made incident on a first spatial light
modulator (SLM1), where the desired quantum states were imprinted on
the signal photon holographically~\cite{Bolduc:2013}. The generated
photonic Kings are subsequently imaged onto a second spatial light
modulator (SLM2) by a $4f$ system.  The second SLM possessing the
desired hologram followed by a single mode optical fibre perform the
projective measurement on the state of the signal photon. Both photons
are sent to avalanche photodiode detectors (APD) and coincidence
counts are recorded by a coincidence box with a coincidence
time window of 3~ns~\cite{Qassim:2014}.

To verify the accurate experimental generation of these states, we
perform quantum state tomography to reconstruct the Husimi $Q$
function, as shown in Fig.~\ref{fig:setup}(b). The average fidelity of
the resulting states is above $90\%$; i.e., $94\%$, $87\%$, $91\%$ and
$93\%$, respectively (see Supplemental Material).

We now study the behaviour of such states under rotations in the
sphere $\mathcal{S}_2$. This is experimentally {realized by}
projective measurements of the Kings  onto themselves after a
rotation $\omega$ around several axes (see
Fig.~\ref{fig:sensitivity}). To demonstrate the high sensitivity to
rotation of these states along arbitrary axes, we perform such
rotations around each axis passing through the Majorana points, and
facets of the Kings constellations. For the cases of $S=3$, 6 and
10, we find four-, five- and three-fold symmetry axes passing
through their Majorana points and three-, three- and five-fold
symmetry axes passing through the normals to the facets of their
constellations, respectively. {Note that, since we are dealing
  with OAM, these rotations correspond to rather abstract mode
  transformations,  although the polar axis still represent a physical
  real-space rotation around the optical axis.}

\begin{figure}[t]
  \centerline{\includegraphics[width=0.95\columnwidth]{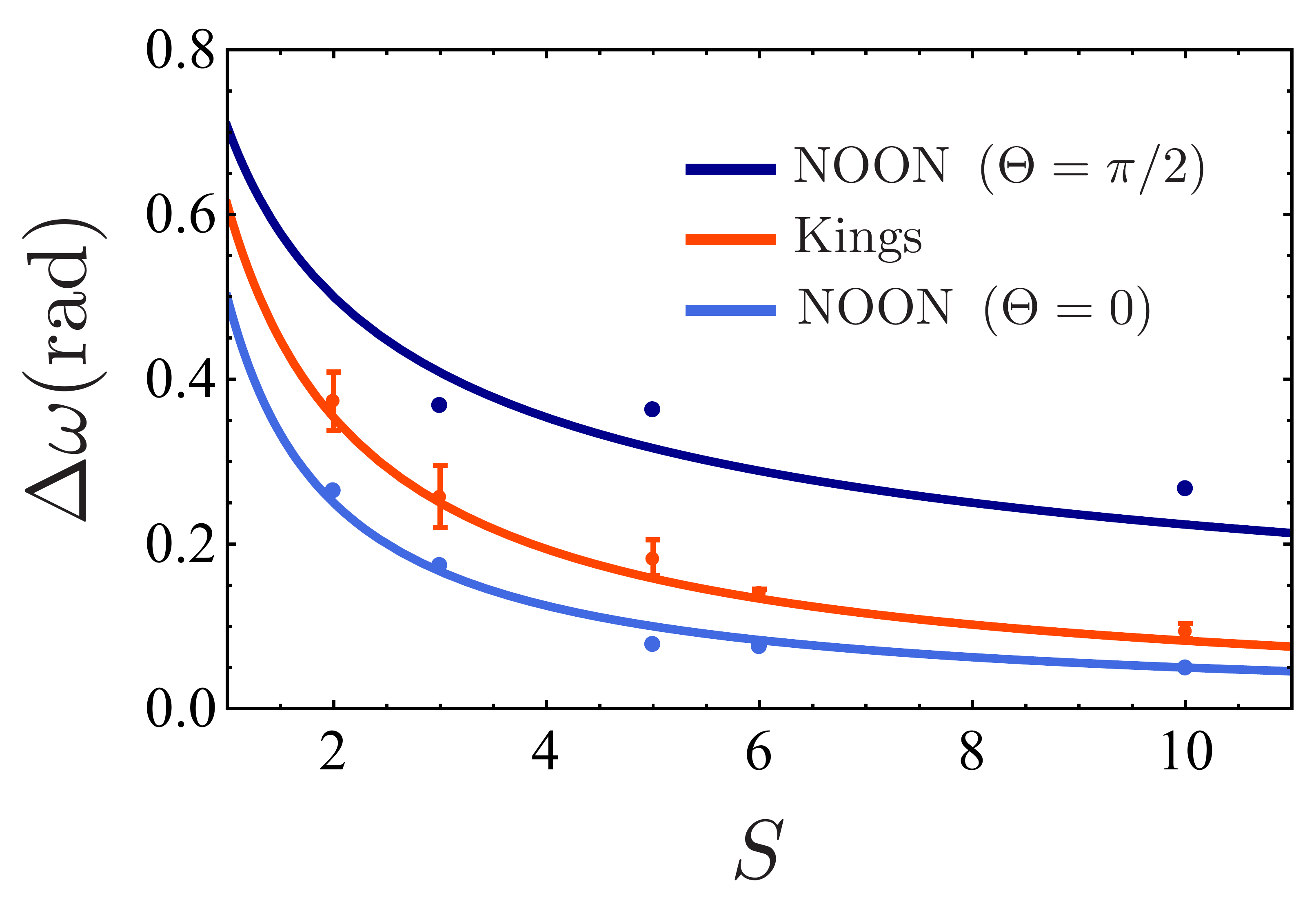}}
  \caption{Rotational sensitivity $\Delta \omega$ for the Kings (red)
    and NOON states (blue). The solid curves correspond to the theory
    predicted in Eq.~(\ref{eq:senKN}) and the points correspond to
    experimental results. In the case of the Kings, we show the mean
    rotational sensitivity over all axes presented in
    Fig.~\ref{fig:sensitivity}, where the error bars correspond to
    variation in sensitivity from axes to axes. For the NOON states,
    we show the rotational sensitivity for rotation axes with
    $\Theta=0$ and $\Theta=\pi/2$.}  \label{fig:noon}
\end{figure}

Finally, in Fig.~\ref{fig:noon} we experimentally check the
sensitivity of the Kings and NOON states. As we can see, the
experimental sensitivity of the Kings is completely independent of the
orientation of the rotation axes (within the error bars). In the limit
of small rotation angles, the NOON states overcome the Kings all the
way up to $\cos \Theta = 1/\sqrt{3}$. Nonetheless, since the Kings
achieve the ideal sensitivity irrespectively of the axis, they are the
most appropriate to detect rotation around arbitrary axes.

The problem of the Kings is closely related to other notions as states
of maximal Wehrl-Lieb entropy~\cite{Baecklund:2014ng}, Platonic
states~\cite{Kolenderski:2008mo}, the Queens of
Quantumness~\cite{Giraud:2010db} or the Thomson
problem~\cite{Thomson:1904qp}. However, there are still many things to
elucidate concerning these links.  They are though a nice illustration of the
connections between different branches of science, and on how some
seemingly simple problems---distributing points in the most symmetric
manner on a sphere---can illuminate such complicated optimization
problems that we have just described.

Thus far, efforts were concentrated in estimating the rotation angle,
which in terms of magnetometry means that we only want to know the
magnetic field magnitude. The Kings will allow for a simultaneous
precise determination of the rotation axis (i.e., the magnetic field
direction). Our experimental results corroborate that this extra
advantage can pave the way to much more refined measurement schemes.

\section*{Funding Information}
  F.~B. acknowledges the support of the Vanier Canada Graduate
  Scholarships Program of the Natural Sciences and Engineering
  Research Council of Canada (NSERC). E.~K. acknowledges the support
  of the Canada Research Chairs (CRC), and Canada Foundation for
  Innovation (CFI) Programs. F.~B., R.~W.~B and E.~K. acknowledge the
  support of the Max Planck--University of Ottawa Centre for Extreme
  and Quantum Photonics.  Z.~H. and J.~R. acknowledge the support from
  the Technology Agency of the Czech Republic (Grant TE01020229), the
  Grant Agency of the Czech Republic (Grant 15-03194S). 
P.~H. and L.~L.~S.~S. acknowledges the support of the
  Spanish MINECO (Grant FIS2015-67963-P).


\end{document}